\documentstyle[./psfig]{./l-aa}

\def\0{\hspace*{0.5em}}

\newcommand{\msun}{\mbox{${\rm M}_\odot$}}
\newcommand{\MO}{\mbox{${\rm M}_\odot$}}

\newcommand{\mpyr}{\mbox{${\rm M}_\odot {\rm yr}^{-1}$}}
\newcommand{\kms}{\mbox{${\rm km~s}^{-1}$}}
\newcommand{\vt}{\mbox{$v_{\rm tr}$}}
\newcommand{\avvt}{\mbox{$\langle v_{\rm tr} \rangle$}}
\newcommand{\Mhe}{\mbox{$M_{\rm He}$}}
\newcommand{\M}{\mbox{$M_\circ$}}
\newcommand{\Mf}{\mbox{$M'_{\rm tot}$}}
\newcommand{\Mfsn}{\mbox{$M''_{\rm tot}$}}
\newcommand{\Mns}{\mbox{$1.4\,{\rm M}_\odot$}}
\newcommand{\Mtot}{\mbox{$M_{\rm tot}$}}
\newcommand{\Pf}{\mbox{$P_f$}}
\newcommand{\Po}{\mbox{$P_\circ$}}
\newcommand{\Af}{\mbox{$a_f$}}
\newcommand{\Ao}{\mbox{$a_\circ$}}
\newcommand{\Mpo}{\mbox{$M_\circ$}}
\newcommand{\Mso}{\mbox{$m_\circ$}}
\newcommand{\Mpf}{\mbox{$M_f$}}
\newcommand{\Msf}{\mbox{$m_f$}}
\newcommand{\qo}{\mbox{$q_\circ$}}
\newcommand{\Vrec}{\mbox{$V_{\rm rec}$}}
\newcommand{\Vorb}{\mbox{$V_{\rm orb}$}}
\newcommand{\Vorbp}{\mbox{$V_{{\rm orb}, 1}$}}
\newcommand{\DMsn}{\mbox{$\Delta M_{\rm sn}$}}
\newcommand{\DMw}{\mbox{$\Delta M_{\rm wind}$}}

\newcommand{\V}{\mbox{\em V}}

\def\apgt{\ {\raise-.5ex\hbox{$\buildrel>\over\sim$}}\ }
\def\aplt{\ {\raise-.5ex\hbox{$\buildrel<\over\sim$}}\ }

\begin{document}


\title{On the origin of the difference between the runaway velocities
of the OB-supergiant X-ray Binaries and the Be/X-ray Binaries}

\author{E.~P.~J.~van~den~Heuvel\inst{1, 3}
\and	Simon~F.~Portegies~Zwart\inst{1, 2}\thanks{Hubble Fellow}
\and	Dipankar Bhattacharya\inst{4}
\and	Lex Kaper\inst{1}}
\offprints {S.~F.~Portegies~Zwart;  e-mail: spz@komodo.bu.edu}

\institute{Astronomical Institute {\em Anton Pannekoek}, 
           Kruislaan 403, 1098 SJ Amsterdam, The Netherlands
\and	Department of Astronomy,
		  Boston University,
		  725 Commonwealth Ave.,
		  Boston, MA 02215, USA 
\and
 	   Institute for Theoretical Physics UCSB, Santa Barbara, 
	   California 93106-4030
\and	
	 Raman Research Institute, 560 080 Bangalore, India
} 

\date{Received; accepted: 04.05.2000}
\maketitle
\markboth{Runaway velocities of X-ray binaries}{E.P.J.~van~den~Heuvel et al.}

\begin{abstract}
The recent finding by Chevalier \& Ilovaisky (1998) from 
{\it Hipparcos} observations that OB-supergiant X-ray binaries have
relatively large runaway velocities (mean peculiar tangential
velocity\footnote{The values given here are not identical (though
similar) to those listed in Chevalier \& Ilovaisky (1998). The
corrections we applied are outlined below.} $\avvt = 42 \pm 14$~\kms),
whereas Be/X-ray binaries have low runaway velocities ($\avvt = 15 \pm
6$\,\kms), provides confirmation of the current models for the
formation of these two types of systems. These predict a difference in
runaway velocity of this order of magnitude.  This difference
basically results from the variation of the fractional helium core
mass as a function of stellar mass, in combination with the
conservation of orbital angular momentum during the mass transfer
phase that preceded the formation of the compact object in the
system. This combination results into: (i) Systematically narrower
pre-supernova orbits in the OB-supergiant systems than in the
Be-systems, and (ii) A larger fractional amount of mass ejected in the
supernovae in high-mass systems relative to systems of lower mass.
Regardless of possible kick velocities imparted to neutron stars at
birth, this combination leads to a considerable difference in average
runaway velocity between these two groups.  If one includes the
possibility for non-conservative mass transfer the predicted
difference between the runaway velocity of the two groups becomes even
more pronounced.  The observed low runaway velocities of the Be/X-ray
binaries confirm that in most cases not more than 1 to 2\,\msun\ was
ejected in the supernovae that produced their neutron stars. This, in
combination with the --on average-- large orbital eccentricities of
these systems, indicates that their neutron stars must have received a
velocity kick in the range 60 - 250 \,\kms\ at birth.  The
considerable runaway velocity of Cygnus X-1 ($\vt = 50\pm15$\, \kms)
shows that also with the formation of a black hole considerable mass
ejection takes place.
\end{abstract}


\section{Introduction}

A high-mass X-ray binary (HMXB) consists of a massive OB-type star and
a compact X-ray source, a neutron star or a black hole. The X-ray
source is powered by accretion of wind material, though in some
systems mass transfer takes place through Roche-lobe overflow; the
compact stars in the latter systems are surrounded by an accretion
disk. Since wind accretion plays an important role, in practice only
an OB supergiant or a Be-star companion have a strong enough stellar
wind to result in observable X-ray emission. In a Be/X-ray binary the
X-ray source is only observed when the neutron star moves through the
dense Be-star disk at periatron passage. About 75\% of the known HMXBs
are Be/X-ray binaries, although this is a lower limit given their
transient character.

Chevalier \& Ilovaisky (1998) derived the proper motions for a sample
of HMXBs from {\it Hipparcos} measurements. The four OB-supergiant
HMXBs for which proper motions are available (0114+65, 0900-40 [Vela
X-1], 1700-37 and Cyg X-1) have relatively large peculiar tangential
velocities. Some corrections to the values given by these authors are
needed (cf.\ Steele et al.\, 1998, Kaper et al. 1999). Taking these into
account (Table 1) the mean peculiar velocity of these systems is $42
\pm 14$\,\kms. It was already known that the OB-supergiant system of
1538-52 (QV Nor) has a peculiar radial velocity of about 90\, \kms\
with respect to its local standard of rest (Crampton et al. 1978; Gies
\& Bolton 1986; van Oijen 1989). For the 13 Be/X-ray binaries with
measured proper motions Chevalier \& Ilovaisky found peculiar
tangential velocities ranging from $\vt = 3.3 \pm 0.7$ to $21 \pm
7.4$\, \kms, with an average of $\avvt = 11.4 \pm 6.6$\,\kms.  Again,
after corrections (see Sect.\,2) and excluding the Oe systems X~Per
(0352+309) and V725~Tau (0535+262), one finds for the genuine Be/X-ray
binary a slightly higher value of $\vt = 15 \pm 6$\,\kms.  

We would like to point out here that these mean values are in good
agreement with the runaway velocities of these two types of systems
predicted on the basis of simple ``conservative'' evolutionary models
(van den Heuvel 1983, 1985, 1994; Habets 1985; van den Heuvel \&
Rappaport 1987) and even better agreement is obtained when mass is not
conserved in the transfer process (Portegies Zwart 2000).  The effect
of sudden mass loss during the supernova explosion is taken into
account and in a massive binary this is the dominant contribution to
the runaway velocity; a random kick velocity of a few hundred \kms\
imparted to the neutron star at birth (see e.g.\ Hartman 1997) has
only a small effect, as the kick's impulse has to be distributed over
the entire massive ($\apgt 15$\,\msun) system. (See Portegies Zwart \&
van den Heuvel 1999, for arguments in favor of kicks). Therefore, in
first-order approximation, these kicks can be neglected in calculating
the runaway velocities of HMXBs, but {\em not} in calculating their
orbital eccentricities (see Sects.\, 3.4 and 3.5).

The aim of the present paper is to give a quantitative assessment of
the above-mentioned conjectures. It should be noted here that five
Be-star systems in the Be/X-ray binary sample studied by Chevalier \&
Ilovaisky (1998) are of spectral type B4~Ve or later (masses $\leq 6
M{_\odot}$). The companions of these stars might be white dwarfs
instead of neutron stars. Therefore, a supernova explosion is not
necessarily the reason for their (excess) space velocity, which, in
any case, is relatively small. It may be due to the typical random
velocities observed in young stellar systems. Leaving these late-type
Be/X-ray binaries out does not result in a significant change in the
observed mean peculiar velocity of the Be-systems.
Furthermore, there is some doubt concerning the use of the
distances based on {\it Hipparcos} parallaxes of several of the other
Be-systems, as these distances differ very much from the distances
determined in other ways, e.g. by using reddening etc. (Steele et
al. 1998). In Sect.\, 2 we therefore critically examine the distances
and proper motions of all the systems with Be companions. 

In Sects.\, 3.1 and 3.2 we present an analytical calculation of the
expected runaway velocities and orbital eccentricities of typical
OB-supergiant and Be HMXBs, on the basis of the standard evolutionary
models for these systems, adopting conservative mass transfer during
phases of mass exchange, and including the effects of stellar-wind
mass loss for the OB-supergiant systems. In Sect.\ 4 we discuss the
effect of non-conservative mass transfer on the runaway velocity and
in Sect.\, 5.1 for the Be/X-ray binaries with known orbital
eccentricities. We calculate which kick velocities should be imparted
to the neutron stars of Be/X-ray binaries in order to produce their,
on average, large orbital eccentricities (since the mass-loss effects
alone cannot produce these). In Sect.\, 5.2, as an alternative, we
compare the observed runaway velocities and orbital eccentricities of
the Be/X-ray binaries with those expected on the basis of symmetric
mass ejection and show that without kicks their combination of high
orbital eccentricities and low space velocities cannot be
explained. Our conclusions are summarized in Sect.\, 6.

\section{The observed peculiar tangential velocities of HMXBs} 

The 4 OB-supergiant systems in the {\it Hipparcos} sample of Chevalier
\& Ilovaisky (1998) have distances larger than 1 kpc, which is too
remote for a reliable parallax determination. For these systems they
estimated the distances based on the spectral type, visual magnitude
and reddening, and eventually the strength and velocity of
interstellar absorption features, etc. After correcting for the
peculiar solar motion and differential galactic rotation (see also
Moffat et al. 1998) the {\it Hipparcos} proper motions result in the
peculiar tangential velocities listed in Table~1. Chevalier \&
Ilovaisky give a mean peculiar tangential velocity of $\vt = 41.5 \pm
15$\,\kms. We derive a similar value of $42 \pm 14$~\kms.

For the Be-systems, Chevalier \& Ilovaisky use the {\it Hipparcos}
parallaxes to determine the distances. For some systems this leads to
very surprising results. In particular, Steele et al. (1998) point out
that for the system of 0236+610 (LSA + $61^{\circ} 303$) the 
{\it Hipparcos} parallax leads to a ten times smaller distance than the
distance estimated from the spectral type and reddening. These authors
convincingly show that for this system the distance estimate based on
the {\it Hipparcos} parallax cannot be correct; the distance of the
system must be of order 1.8~kpc instead of the 177~pc determined from
the {\it Hipparcos} parallax. Similarly, from a variety of criteria
they find that for A0535+262 the distance must be $>1.3$~kpc, instead
of the 300~pc determined from the {\it Hipparcos} parallax. Steele et
al. point out that for both systems the {\it Hipparcos} parallaxes are
smaller than 3 times their probable (measurement) error, and are
therefore not reliable. In such a case one cannot reliably use the
{\it Hipparcos} parallax to determine the distance. 
With the {\it Hipparcos} distances the OB-star companions of 0236+610
and 0535+262 would become highly underluminous for their spectral
types, and would be very peculiar stars, as was already noticed by
Chevalier \& Ilovaisky (1998). On the other hand, using alternative
distance criteria, their absolute luminosities become perfectly normal
for their spectral types. This gives confidence that the latter
distances are more reliable.

The systems including a Be star with spectral type later than B4~V
(mass $\leq 6 M_{\odot}$) may well have white dwarfs instead of
neutron stars as companions (Portegies Zwart 1995). Therefore, their
space velocity is not necessarily caused by a supernova explosion,
which is the scenario we exploit in this paper. Excluding these
systems, the observed mean peculiar velocity hardly changes ($\langle
\vt \rangle = 14.4 \pm 6.6$~\kms\ in stead of $15 \pm 6$, excluding
X~Per and V725~Tau), and since the nature of their compact companions
is not known anyway (e.g. no X-ray pulsations observed which would
identify the compact star as a neutron star), we decided to leave them
in the calculation of the mean peculiar velocity. The peculiar
tangential velocities of X~Per (0352+309, O9~III-IVe, 27~\kms) and
V725~Tau (0535+262, O9.7~IIe, 97~\kms) are relatively high; their
early spectral types suggest that they have masses comparable to those
of the OB supergiants, so that, like the OB-supergiant systems, they
would also originate from relatively massive binary systems. In
Table~1 we list the peculiar tangential velocity for each individual
system and calculate the average for different subsamples. We left out
the Be~star $\gamma$~Cas, because its X-ray binary nature is not
clear; furthermore, its X-ray spectrum is consistent with that of a
white dwarf (Haberl 1995).

For the systems 0236+610, 0535+262, 1036-565 and 1145-619 the 
{\it Hipparcos} parallaxes yield absolute visual magnitudes very different
from those expected on the basis of the OB-spectral types of the
stars. In these cases, the {\it Hipparcos} parallax measurements are
less than three times their probable errors and thus not reliable. For
these stars we therefore used the distances determined from spectral
type and reddening, which yield absolute visual magnitudes consistent
with their spectral types. 

We rederived the peculiar tangential velocities relative to the local
restframe from the {\it Hipparcos} proper motions (cf.\ Kaper et
al. 1999). Table~1 lists the peculiar tangential velocity corrected
for the peculiar solar motion and differential galactic rotation for
three different distances ($d/1.4$, $d$, $1.4d$, following Gies \&
Bolton 1986). The uncertainty in distance (and thus in peculiar
motion) is difficult to estimate; therefore, we calculated the space
velocity for different values of the distance. The peculiar tangential
velocities for the HMXBs discussed in Clark \& Dolan (1999) are
identical to ours for the OB-supergiant systems, though they find
different values for the Be/X-ray binaries X~Per ($15 \pm 3$~\kms,
$d=700$~pc), V725~Tau ($57 \pm 14$~\kms, $d=2$~kpc), and 1145-619 ($17
\pm 7$~\kms, $d=510$~pc).  Obviously, the precise values for the
peculiar motion depend on the adopted model for the galactic rotation;
we used the formalism employed in Comer\'{o}n et al. (1998). For the
OB-supergiant systems in the sample of Chevalier \& Ilovaisky (1998)
also the radial velocities are available from literature. This is not
the case for the Be/X-ray binary systems. Therefore, we only consider
the two components of the tangential velocity for the comparison of
the kinematic properties of the two groups. The table shows that,
leaving the two O-emission systems out, the Be/X-ray binaries have low
space velocities: $15 \pm 6$~\kms.

\begin{table*}
\caption[]{Peculiar transverse velocities (with respect to the local
standard of rest) of the HMXB sample of Chevalier \& Ilovaisky (1998),
after correcting the distances of the four Be-systems for which the
{\it Hipparcos} parallaxes are not significant (see text). For all systems
we recalculated the corrections for solar motion and Galactic
rotation.  }
\begin{tabular}{lllllll}
X-ray source & Name & Spectral type & Distance $d$ & 
\multicolumn{3}{c}{$\vt$ [\kms]}  \\  
  & & & [kpc] & $d/1.4$ & $d$ & $1.4d$  \\ 
\hline
 0114+650 & V662Cas  & B0.5\,Ib  & 3.8 & 14.2 & 26.4 & \043.6 \\
 0900-403 & GP Vel   & B0.5\,Ib  & 1.8 & 22.7 & 34.0 & \050.0 \\
 1700-377 & V884Sco  & O6.5\,If+ & 1.7 & 46.0 & 58.0 & \074.8 \\
 1956+350 & V1357Cyg & O9.7\,Iab & 2.5 & 35.3 & 47.4 & \064.5 \\
\hline
 0236+610 & LSI+$61^{\circ}$303 & B0\,IIIe   & 1.8  & \02.6 & \09.0 &
\019.4 \\
 0352+309 & X Per               & O9\,III-IVe& 0.8  & 23.6 & 27.3 &
\033.4 \\
 0521+373 & HD34921             & B0\,IVpe   & 1.05 & 18.4 & 23.4 &
\032.1  \\
 0535+262 & V725 Tau            & O9.7\,IIe  & 2.0  & 73.2 & 96.5 &
129.1  \\ 
 0739-529 & HD63666             & B7\,IV-Ve  & 0.52  & \09.7 & 12.2 &
\016.9  \\
 0749-600 & HD65663             & B8\,IIIe   & 0.4  & \06.9 & \09.9 &
\016.0  \\
 1036-565 & HD91188             & B4\,IIIe   & 0.5  & 18.3 & 20.4 &
\023.4  \\
 1145-619 & V801 Cen            & B1\,Ve     & 1.1  & \07.9 & \05.8 &
\0\06.1  \\
 1249-637 & BZ Cru              & B0\,IIIe   & 0.3  & 12.9 & 13.6 &
\016.3  \\ 
 1253-761 & HD109857            & B7\,Ve     & 0.24 & 17.6 & 23.7 &
\033.8  \\
 1255-567 & $\mu^{2}$ Cru       & B5\,Ve     & 0.11 & 12.5 & 13.5 &
\016.9  \\ 
\hline
\end{tabular}
\end{table*}

\section{Runaway velocities expected on the basis of models with
conservative mass transfer and symmetric mass ejection.}

\subsection{Change of orbital period due to mass transfer}

We only consider here so-called case B mass transfer since for the
evolution of massive close binaries this is the dominant mode of mass
transfer (cf.\ Paczynski 1971; van den Heuvel 1994, but see Wellstein
\& Langer 1999). In case B the mass transfer starts after the primary
has terminated core-hydrogen burning, and before core-helium
ignition. After the mass transfer in this case the remnant of the
primary star is its helium core, while its entire hydrogen-rich
envelope has been transferred to the secondary, which due to this
became the more massive component of the system.  There is a simple
relation between the mass of the helium core \Mhe\ and that of its
progenitor \M (see for example van der Linden 1982; Iben \& Tutukov
1985).  We adopt here the relation given by Iben \& Tutukov (1985):

\begin{equation}
                         \Mhe  = 0.058 \M^{1.57},  
\label{Eq:1}\end{equation}                         
which results in a fractional helium core mass $p$ given by:
\begin{equation}
                         p = \Mhe/\M = 0.058\M^{0.57}.
\label{Eq:2}\end{equation}

The change in orbital period of the system in case of conservative
mass transfer (i.e.: conservation of total system mass \Mtot\ and orbital
angular momentum $J$) and initially circular orbits is (Paczynski 1971;
van den Heuvel 1994):
\begin{equation}
        {\Pf \over \Po} = \left( {\Mpo \Mso \over \Mpf \Msf}
                          \right)^3, 
\label{Eq:3}\end{equation} 
where \Po, \Mpo\ and $\Mso = \Mtot - \Mpo$ denote the orbital period
and component masses before the mass transfer, and \Pf, \Mpf\ and
$\Msf = \Mtot-\Mpf$\, are the orbital period and component masses after
the transfer. The transformation between orbital separation and the
orbital period is given by Kepler's third law.

Introducing the initial mass ratio $\qo = \Mso/\Mpo$ and using
equation~\ref{Eq:2}, equation~\ref{Eq:3} can be written as
\begin{equation}
	{\Pf \over \Po} = \left( \frac {\qo} {p(q_{o}+1-p)} \right)^3.
\label{Eq:4}\end{equation}
                         
Since according to equation~\ref{Eq:2}, $p$ increases for increasing
stellar mass, one observes that, due to the third power in
equation~\ref{Eq:4}, for the same \qo\ the orbital period of a very
massive system increases much less as a result of the mass transfer,
than for systems of lower mass. (The term $(q_{o} +1-p)$ changes much
less than $p$ itself for increasing stellar mass, so for a given
$q_{o}$ this term has only a modest effect.)  This is the main reason
for the systematically longer orbital periods of the Be/X-ray binaries
(always $> 16$\,days) relative to those of the OB-supergiant HMXBs (in
all but one case: between 1.4 days and 11 days, cf.\ van den Heuvel,
1983, 1985, 1994). This is illustrated in Table\,\ref{Tab:model} where
we list the relative post-mass-transfer periods $\Pf/\Po$ for typical
Be/X-ray binary progenitor systems, with $\Mpo = 10$\,\msun\ and
12\,\msun, respectively, and for two typical OB-supergiant HMXB
progenitors with $\Mpo = 25$\,\MO\ and 35\,\MO, respectively, for \qo\
values ranging from 0.4 through 0.8.

\subsection{Possible effects of further mass transfer and stellar
winds on the orbits} 

\subsubsection{Case BB mass transfer} 

The helium cores left by the 10\,\MO\ and the 12\,\MO\ stars have
masses of 2.15\,\MO\ and 2.87\,\MO, respectively. During helium-shell
burning, when these stars have CO-cores, their outer layers may expand
to dimensions of a few to several tens of solar radii, and a second,
so-called Case BB, mass transfer may ensue before their cores
collapse to neutron stars (Habets 1985,1986ab). However, since the
radius of the 2.87\,\MO\ helium star will not exceed 5~$R_{\odot}$, a
second mass transfer phase is unlikely to occur here. In the case of
the 2.15\,\MO\ helium star, which does attain a large radius, the
amount of mass that is in the extended envelope is not more than
0.1\,\MO. For these reasons, we will neglect here the effects of Case
BB mass transfer, and will assume that these helium stars do not lose
any mass before their final supernova explosion. This means, that we
will slightly overestimate the imparted runaway velocities (as the
orbits at the time of the explosion will be slightly wider than we
assume, and the ejected amounts of mass will be somewhat smaller than
we assume).

\subsubsection{Stellar-wind mass loss in massive stars}

Since wind mass-loss rates from Wolf-Rayet (WR) stars --massive helium
stars-- are much larger (viz.: $\apgt 10^{-5}$\,\mpyr) than the
mass-loss rates of lower-mass main-sequence stars, for the sake of
argument (in order to include only the largest effects) we take
into account the effects of the wind mass loss during the WR phase.
The effects of these winds are: (1) to widen the orbits, and
(2) to considerably decrease the mass of the helium ($\equiv$\, WR)
star before its core collapses.  In the cases of the 25\,\MO\ and
35\,\MO\ primary stars, the masses of the helium cores are 9.1\,\msun\
and 15.4\,\MO, respectively. Such stars live $9 \times 10^5$ years and
$7.5 \times 10^5$ years, respectively, and are expected to lose about
4.0\,\msun\ and 7.4\,\MO\ through their wind during this phase of
their evolution, respectively\footnote{We assumed here wind mass-loss
rates of $0.5 \times 10^{-5}$\,\mpyr\ for the 9.1\,\MO\ star and
$10^{-5}$\,\mpyr\ for the 15.4\,\MO\ star, respectively. These
rates are in good agreement with observed WR-wind mass-loss rates
(cf.\ Leitherer et al. 1995), but are lower than the rates adopted by
Woosley et al.\ (1995), which {\bf may} overestimate the real
mass-loss rates, since they give for all initial helium star masses,
final masses before core collapse of only about 4\,\MO.}

Thus, at the moment of the supernova explosion the collapsing cores of
these stars will have masses of 5\,\MO\ and 8.0\,\MO, respectively.
To keep the same notation we will express the relative mass loss in
the stellar wind with $\delta = \Delta M_{\rm wind} / \Mpo$.  The
value for $\delta$ is 0.16 for a primary with a mass of 25\,\msun\ and
0.21 for a 35\,\msun\ primary star. In the cases of no wind mass loss
(in lower mass primaries): $\delta = 0$.

The wind mass loss will change the post-mass-transfer orbits as
follows (van
den Heuvel 1994):
\begin{equation}
                          d \log a  = - d \log \Mtot,
\label{Eq:5}
\end{equation}
and
\begin{equation}
                          d \log P  = - 2d \log \Mtot,
\label{Eq:6}
\end{equation}

where $a$ is the orbital separation and $\Mtot$ the total system mass. 

Eq.~(\ref{Eq:6}) results in:
\begin{equation}
                          P/P'  = \left( {\Mf \over \Mtot} \right)^2
\, ,
\label{Eq:7}
\end{equation}
where $P$ and $P'$ correspond to the system masses $\Mtot$ and \Mf,
respectively. $\Mtot$ is the total mass at the beginning of the WR
phase and \Mf\ the total system mass at the end of this phase, just
prior to the supernova explosion of the WR Star.  The orbital
separation after mass transfer and additional WR mass loss phase is
expressed as:
\begin{equation} 
	{a' \over \Ao} =  {\Af \over \Ao} {a' \over \Af} 	
			\equiv \left( {\qo \over p(1+\qo-p)}  \right)^2
 			       \left(1 - {\delta \over 1+\qo} \right)^{-1}
\label{Eq:8}
\end{equation} 

\subsection{Runaway velocities induced by symmetric supernova mass ejection}

The runaway velocity imparted to the system by the supernova mass loss
is calculated from the loss of momentum of the system during the
explosion: $-\Vorbp \DMsn$, where \Vorbp\ is the orbital velocity of
the helium star prior to the explosion and \DMsn\, is the amount of mass
ejected in the supernova.

We assume all compact remnants to be a \Mns\ neutron star. Then
$\Delta M_{\rm sn}$ is given by $\Delta M_{\rm sn} = (p-\delta) M_{o} -1.4
M_{\odot}$.  The remaining mass of the system is:
\begin{equation}
            \Mfsn = \Msf + 1.4\MO =  (q_{o} +1-p) M_{o} + 1.4 M_{\odot}.
\label{Eq:10}
\end{equation}

This yields a recoil velocity (or runaway velocity) of the system of:
\begin{eqnarray}
      \Vrec & = & V_{\rm orb,1} 
	          {\DMsn \over (q_{o}+1-p) M_{o} + 1.4 M_{\odot}}
\label{Eq:11}
\end{eqnarray}
Here the second term in the right argument is simply the post
supernova eccentricity and we may write simply
\begin{equation}
      \Vrec  =  e V_{\rm orb,1} 
\label{Eq:11b}
\end{equation}

The relative orbital velocity before the explosion is $\sqrt{G
\Mf/a'}$. One therefore has:
\begin{equation}
V_{\rm orb,1} = \left( {GM_{o} \over a_{o}} \right)^{1/2}  
	        { p(1+q_{o}-p)^{2} \over q_{o}(1+q_{o})^{1/2}}
\label{Eq:12}
\end{equation}

Substitution of Eq.~(11) into Eq.~(10) results now in 
\begin{eqnarray}
\V_{\rm rec} &=& \left({GM_{o} \over a_{o}}\right)^{1/2}  \nonumber \\
	     &\times& {p(1+q_{o}-p)^{2} \over q_{o}(1+q_{o})^{1/2}} \;
                {(p-\delta - 1.4M_{\odot}/M_{o}) \over (q_{o} +1 -p+1.4
			M_{\odot}/M_{o})} 
\label{Eq:14}\end{eqnarray}
which in numerical form becomes:

\begin{eqnarray}
V_{\rm rec} &=& 212.9 [\kms] \left( {M_{o} \over [M_{\odot}]}
                                    {[{\rm days}] \over P_{o}} 
                           \right)^{1/3} \nonumber \\
            &\times&  {p(q_{o}+1-p)^{2} 
				(p- \delta - 1.4M_{\odot}/M_{o})
				(1+q_{o}-\delta) \over q_{o}(1+q_{o}
				)^{2/3} (q_{o} +1-p+1.4M_{\odot}/M_{o}) } 
\label{Eq:15}\end{eqnarray}


\begin{figure}[t]
\centerline{
\psfig{file=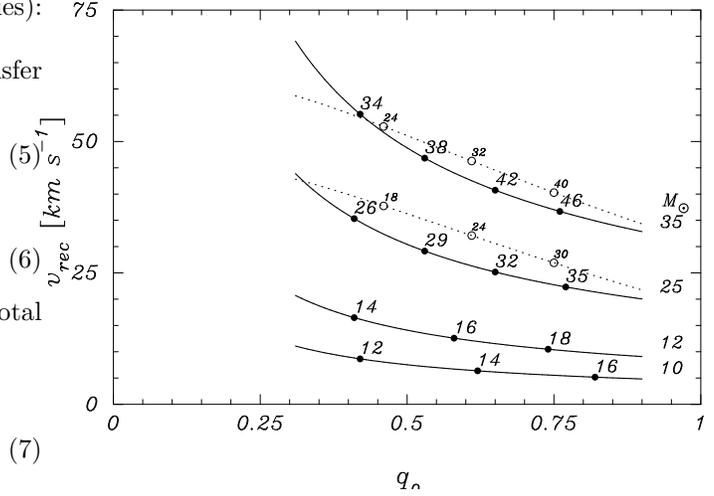,height=7.5cm,angle=-90}}
\caption[]{The recoil velocity of X-ray binaries induced by the
supernova as a function of the initial mass ratio.  The initial
orbital period is 5~days for all binaries.  From bottom to top the
solid lines represent the recoil velocities for binaries with an
initial primary mass of 10\,\msun, 12~\msun, 25~\msun\, and 35~\msun,
respectively.  The numbers printed above $\bullet$ on the curves
indicate the mass of the runaway star in \msun.  Mass loss in the
Wolf-Rayet phase is taken into account for the top two curves
(25~\msun\, and 35~\msun\, zero-age primaries).  The dotted lines show
the same evolutionary calculations but relaxing the assumption that
mass is conserved during transfer (see sect.\,\ref{non_conservative}).
}
\label{fig:conservative}
\end{figure}

Fig.~1 shows for $\Po = 5$\,days the values of \Vrec\ as a function
of \qo\ for the four primary masses of Table~2, using the \DMsn\ and
\DMw\ as given above. The figure shows that for the ``Be-systems''
(initial primary masses 10\,\MO\ and 12\,\MO\ yielding Be-star masses
ranging from 11.85\,\MO\ to 18.7\,\MO) the expected recoil velocities
range from 5 to 21\, \kms, whereas for the ``OB-supergiant systems''
(with OB-companions between 25\,\MO\ and 40\,\MO) they range between
21 and $> 80$\,\kms, respectively. These velocities correspond to
transverse velocities that are $\pi/4$ times these values, i.e.: 3.9
to 17 \kms\, for the Be-systems, and 16.5 to $>71$ \kms\, for the
OB-supergiant systems with neutron stars. Thus one expects average
transverse velocities of order 10.5~\kms\ and 45~\kms\ for the
Be/X-ray binaries and OB-supergiant systems, respectively.  For both
the Be/X-ray binaries and the OB-supergiant systems Be/X-ray the
predicted and observed mean transverse runaway velocities agree 
well: $15 \pm 6$~\kms and $42 \pm 14$, respectively.


As Eq.~(\ref{Eq:15}) shows, the dependence of the recoil velocity
on \Po\ is rather weak, so for initial orbital periods between a few
days and 10\,days these results don't change by more than a factor
1.5. Therefore, certainly qualitatively, Fig.~1 is representative
for the two types of systems.  Eq.~(\ref{Eq:15}) further shows that
the large difference in runaway velocity between the two types of
systems is due to a combination of two factors, as follows: (1) the
larger fractional helium core masses ($p$) in the more massive
systems, which cause their pre-supernova orbital periods to be shorter
and thus their pre-supernova orbital velocities to be larger than
those of the lower-mass systems; and (2) the much lower amounts of
mass ejected ($\Delta M_{\rm sn}$) in the lower mass systems compared to
the systems of higher mass, which leads to a lower recoil effect.

Relaxing the assumption that mass is conserved during the phase of
mass transfer changes little, which we will discuss now.  

\begin{table*}
\caption[]{ Resulting orbital parameters and runaway velocities for a
number of characteristic initial primary masses (first column) and
mass ratios (second column). The increase in orbital period
$P_{f}/P_{o}$ due to the mass transfer and the mass of the
OB-component of the resulting X-ray binary are listed in columns 3 and
4, respectively. Columns 5, 6, 7, 8 and 9 list for the case of
symmetric supernova mass ejection: the resulting runaway velocity of
the system $V_{\rm rec}$ (for an assumed initial orbital period
$P_{o}=5^{d}$), $V_{\rm rec}$ expressed as a fraction of the relative
orbital velocity $V_{o}$ in the initial orbit, the orbital
eccentricity, the post-supernova orbital period $P_{f}''/P_{o}$, and
the runaway velocity in the postsupernova system, $f_{v}$, as defined
by equation (22), respectively.}

\begin{minipage}{8cm}
\renewcommand{\footnoterule}{}
\begin{tabular}{llrlrccrc}
\hline\noalign{\smallskip}
\multicolumn{1}{l}{\hspace{-2mm}\Mpo}         & 
\multicolumn{1}{c}{\qo}                       & 
\multicolumn{1}{c}{${P_{f} \over P_{o}}$}     &  
\multicolumn{1}{c}{\Msf}                      & 
\multicolumn{1}{c}{\hspace{-4mm}$V_{\rm rec}$\footnote{for $P_{o}=5^{\rm d}$}}             & 
\multicolumn{1}{c}{${\Vrec/V_{\rm \circ}}$}     & 
\multicolumn{1}{c}{$e$}                       & 
\multicolumn{1}{c}{${P_{f}''/\Po}$}     &  
\multicolumn{1}{c}{$f_v$}                     \\ 
\noalign{\smallskip}
\multicolumn{1}{l}{\hspace{-2mm}[\msun]}      &
\multicolumn{1}{c}{ }                         &
\multicolumn{1}{c}{ }                         &
\multicolumn{1}{c}{[\msun]}                   & 
\multicolumn{1}{c}{\hspace{-4mm} [\kms]}       &
\multicolumn{1}{c}{ }                         &
\multicolumn{1}{c}{ }                         &
\multicolumn{1}{c}{ }                         &
\multicolumn{1}{c}{ }                         \\
%
\noalign{\smallskip}\hline\noalign{\smallskip}
10      & 0.8 & 12.93 & 15.8 &  5.33 & 0.022 & 0.045 & 14.2 & 0.045 \\
        & 0.6 &  8.18 & 13.8 &  6.64 & 0.027 & 0.051 &  9.1 & 0.050 \\
        & 0.4 &  3.86 & 11.8 &  9.18 & 0.036 & 0.060 &  4.4 & 0.060 \\
        & 0.3 &  2.13 & 10.8 & 11.64 & 0.046 & 0.065 &  2.4 & 0.065 \\
12      & 0.8 &  9.84 & 18.7 & 10.20 & 0.040 & 0.073 & 11.4 & 0.073 \\
        & 0.6 &  6.26 & 16.3 & 12.70 & 0.048 & 0.083 &  7.4 & 0.083 \\
        & 0.4 &  2.99 & 13.9 & 17.55 & 0.065 & 0.096 &  3.7 & 0.096 \\
        & 0.3 &  1.65 & 10.7 & 22.25 & 0.082 & 0.104 &  2.0 & 0.104 \\
25      & 0.8 &  3.61 & 35.9 & 22.70 & 0.069 & 0.098 &  5.3 & 0.098 \\
        & 0.6 &  2.38 & 30.9 & 28.01 & 0.083 & 0.114 &  3.8 & 0.115 \\
        & 0.4 &  1.20 & 25.9 & 38.23 & 0.111 & 0.135 &  2.0 & 0.136 \\
        & 0.3 &  0.69 & 23.4 & 48.07 & 0.140 & 0.148 &  1.2 & 0.149 \\
35      & 0.8 &  2.39 & 47.6 & 38.00 & 0.112 & 0.136 &  4.1 & 0.137 \\
        & 0.6 &  1.62 & 40.6 & 46.52 & 0.139 & 0.158 &  2.9 & 0.160 \\
        & 0.4 &  0.85 & 33.6 & 62.76 & 0.201 & 0.190 &  1.8 & 0.193 \\
        & 0.3 &  0.50 & 30.1 & 78.29 & 0.250 & 0.211 &  1.2 & 0.213 \\
\hline
\end{tabular}
\label{Tab:model}
\end{minipage}
\end{table*}

\section{The effects of non-conservative mass transfer}\label{non_conservative}

In the above it was assumed that the case B mass transfer was
conservative in all systems. For the Be-systems this ``conservative''
assumption seems confirmed quite straightforwardly as the Be nature is
interpreted by the accretion of angular momentum and thus of mass. On
the other hand, for the OB-supergiant X-ray binaries several authors
(starting with Flannery and Ulrich 1977 for the Cen X-3
system) have pointed out that certainly in part of the systems the
mass transfer has been non-conservative, and there is a considerable
evolution for massive close binaries altogether (De Loore \& De Greeve
1992). Indeed, close Wolf-Rayet binaries with high mass ratios $q =
M_{\rm WR}/M_{OB}$ such as CQ Cep ($P=1.64$ days, $q=1.19$) and CX Cep
($P=2.22$ days, $q=.44$) cannot have been produced by conservative
evolution, and just these systems are the progenitors of the
OB-supergiant X-ray binaries (cf.\, van den Heuvel 1994).

The amount of mass lost from the system during the transfer will
depend on the initial mass ratio \qo\, of the system. For small \qo\,
the companion will accrete little and most of the envelope mass of the
primary will be lost from the system. On the other hand, for large
\qo\, little mass will be lost from the system. Therefore, in order to
study the effect of mass and angular momentum loss on the runaway
velocity, we assume as a first approximation that the fraction
$f$ of the primary's envelope which is accreted by the companion star
is proportional to the initial mass ratio \qo\, (Portegies
Zwart 1995)
\begin{equation}
	f = \qo.
\end{equation}
After mass transfer the secondary mass then becomes
\begin{equation}
	\Msf = \Mpo + f \Mpo (1-p) 
		\equiv \Mpo \qo (2-p).
\end{equation}

The gas lost by the donor leaves with low velocity but gains angular
momentum via the interaction with the companion star.  It finally
leaves the binary system via the second Lagrangian point $L_2$,
carrying specific angular momentum with it (de Loore \& De Greve 1992).
The specific angular momentum of this lost matter is considerably
larger than what is lost in the specific amount of angular momentum in
the stellar wind (given by Eq.\,\ref{Eq:5}), see for example Soberman
et al.\, (1997).



We assume that the mass that leaves the system carries a fraction
$\beta$ of the specific angular momentum of the binary.  We can then
write the change in orbital separation due to mass transfer as
\begin{equation}	
      {a' \over \Ao} = 
     	 \left( {\Mpf \Msf \over \Mpo \Mso} \right)^{-2}
     	\left( {\Mpf+\Msf \over \Mpo+\Mso} \right)^{2\beta +1}.
\label{Eq:12b}\end{equation} 
and use it as an alternative for Eq.~(\ref{Eq:7}).
Following  Portegies Zwart (1995) we use $\beta = 3$.

Eq.\,(\ref{Eq:8}) then becomes:
\begin{equation} 
	{a' \over \Ao} = \left( {1 \over \qo p(2-p)} \right)^2
			 \left( {1+\qo \over p + \qo(2-p)}
			 \right)^{-2\beta -1}
	      	         \left(1 - {\delta \over 1+\qo} \right)^{-1}
\label{Eq:12c}\end{equation} 

The result of this calculation is presented as the dotted lines in
Fig.~1.  The small number near each $\circ$ indicates the mass of the
visible component, which is smaller than if mass transfer would
proceed conservatively.  One observes that for the same mass of the
visible component of the binary, the runaway velocity of the OB-system
is between 50 and 100 per-cent larger than in the conservative case.
The higher velocity of the binary is mainly caused by the smaller
orbital separation at the moment of the supernova.  We thus see, from
this simple numerical experiment, that non-conservative mass transfer
makes the difference in runaway velocity between the two types of high
mass X-ray binaries considerably larger.

\section{Predicted and observed orbital eccentricities of the
Be/X-ray binaries: evidence for kicks} 

\subsection{Orbital eccentricities of Be/X-ray binaries in case of 	
	    symmetric ejection}

In the case of spherically symmetric mass ejection the orbital
eccentricity induced by the mass loss is (cf.\ Hills 1983):
\begin{equation}
         e =      \frac{\DMsn}{\Mf - \DMsn} 
	  \equiv  \frac{\DMsn}{\Mfsn}.
\label{Eq:13}\end{equation}
One expects that because of the extensive mass transfer and the fact
that before the mass transfer the primary was a (sub)giant, the orbits
just prior to the explosions are circular.  Hence, in case of
spherically symmetric mass ejection, one expects the eccentricities of
the Be/X-ray binaries simply to be given by Eq.~(19).

Table 3 shows that for the Be/X-ray binaries resulting from systems
with an initial primary mass of 10\,\MO, the orbital eccentricities
expected on the basis of Eq.~(\ref{Eq:13}) range from 0.045 to
0.060. For systems resulting from binaries with primaries of 12\,\MO,
the eccentricities range from 0.073 to 0.096. It should be noted
that these are, in fact, overestimates, since we ignored case BB mass
transfer, which would still have somewhat reduced these values.

Orbital eccentricities are known for only five Be/X-ray binaries, as is
listed in Table~3. They range from 0.3 to $> 0.7$, with an average of
about 0.5. For the two long-period systems, 1145-619 and 1258-613, the
orbital eccentricities have not yet been measured, but a rough
estimate of their values can be made as follows. Both systems are
recurrent transients, with outbursts occurring once per orbit, when
the Be star is active, presumably when the stars are near
periastron. The same is true for the systems 0115+63, V0331+53 and
EX02030+375 when their Be-stars are in an active phase. For the latter
systems one calculates from their orbital periods and eccentricities
that within 20 percent their periastron distances are the
same. Apparently, this is the periastron distance required for
triggering an outburst when the Be-star is in an active phase. It thus
seems reasonable to assume that the same is true for 1145-619 and
1258-613. Using this, one finds the latter systems to have
eccentricities of between 0.75 and 0.83, and between 0.70 and 0.80,
respectively. To be conservative, we have indicated this in Table 3
as: $e \geq 0.70$.

Two more systems consisting of a B-star and a neutron star are known:
the binary radio pulsars PSRJ 1259-63 and PSRJ 0045-7319. These have
very eccentric orbits as indicated in Table~3. So, in total we have
nine B-star plus neutron star systems with measured or estimated
orbital eccentricities. References to the orbital parameters of these
systems are indicated in the table.

Observations show that all binaries --including detached ones-- with
orbital periods shorter than 10 days have circular orbits, whereas
detached systems with longer orbital periods do not. This suggests
that in systems with orbital periods shorter than 10 days tidal
forces are effective in circularizing the orbits on a timescale
considerably shorter than the lifetimes of the components of the
binary, whereas in wider systems they apparently are not.  Since the
Be/X-ray binaries are detached systems (cf.\ van den Heuvel \&
Rappaport 1987; van den Heuvel 1994) and have orbital periods longer
than 16 days, it is not surprising that their orbits have not yet been
circularized.

The lifetime of a Be/X-ray binary is expected to be of the order of a
few million years up to about 10\,Myr, the lifetime of the Be
companion of the neutron star. The timescale for tidal circularization
for main-sequence binaries with orbital periods $> 16$ days is at
least a few tens of Myr (see Zahn 1977, Kochanek 1992). Therefore it
is unlikely to catch the binary in the circularization
process. Therefore, we expect that the eccentricities for the Be/X-ray
binaries in Table~3 are still close to those just after the supernova
explosion.  The orbits of the high-mass X-ray binaries with orbital
periods $< 10$ days are all practically circularized by tidal effects.

It should be noted that if the eccentricities of the Be-systems had
resulted from spherically symmetric supernova mass ejection, the
amounts of mass ejected in their supernovae should have been very
large, of order 4 to over 7 solar masses (see for example Iben \&
Tutukov 1998). Since in the case of symmetric mass ejection the
orbital eccentricity and runaway velocity are directly proportional to
each other (see Eqs. [10] and [14]), also the induced runaway
velocities should have been much larger than observed.  For example,
induction of an eccentricity 0.5 with a symmetric explosion requires
1/3 of the system mass to be ejected in the explosion.  With a Be star
of 12\,\MO, as is representative for a typical B0.5~Ve star, and a
neutron star mass of 1.4\,\MO, the initial system mass in this case
must have been 20.1\,\MO, implying an ejected amount of mass of
6.7\MO. In order to obtain a post-supernova orbital period of about
30\,days, as is typical for many Be/X-ray binaries, the initial
orbital period in this example must have been around 11\,days. With
this initial period, and 6.7\,\MO\ explosively ejected, the induced
runaway velocity \Vrec\ would be $\simeq 87$\,\kms (see the equations
in Sect.\, 3.3), which is advariant with the observed velocities.

Similarly, if the induced eccentricity would be 0.3, one finds that
for the same final system mass and orbital period, the runaway
velocity induced by the explosion would have been about 45\,\kms.

As these velocities are some 5, respectively 2.5 times larger than the
mean excess space velocity of 19\,\kms\ [$(4/\pi) \times 15\,\kms$] of
the Be/X-ray binaries, it is clear that the orbital eccentricities of
the Be/X-ray binaries cannot be due purely to symmetric mass ejection
in the supernova explosion.

The only way to obtain both a low runaway velocity of the system and
the high orbital eccentricities listed in Table~3 is by having a small
amount of mass ejected in the supernova, in combination with a
velocity kick of order 60 to 250\,\kms\ imparted to the neutron star
at birth. We describe below how these required kick velocities were
calculated. The randomly directed kick hardly changes the runaway
velocity of the system, as the impulse of the kick imparted to the
neutron star is shared by the entire system (with a mass of order 15
solar masses in the case of the Be/X-ray binaries), and thus the kick
velocity is ``diluted'' to an extra velocity of the system of only 4
to 16\,\kms, in a random direction.  Adding this velocity
quadratically (because of its random direction) to the velocity of
between 5 and 21\,\kms\ imparted to the systems purely by the mass
loss (Fig. 1), one obtains mean runaway velocities of between 6 and
21\,\kms\ for a 60\,\kms\ kick and between 17 and 26\,\kms\ for a
250\,\kms\ kick.

These values are in good agreement with the observed mean excess space
velocities of Be/X-ray binaries of $19 \pm 8$\,\kms ($\pi/4$ times
their average peculiar tangential velocities).

We calculated the minimum kick velocities that have to be imparted to
the neutron star during the supernova explosion in order to obtain the
presently observed orbital eccentricities of the Be-systems in
Table~3. We used the equations derived by Wijers et al. (1992). The
minimum required kick velocitiy is the one that is imparted in the
orbital plane in the direction of motion of the pre-supernova star
(assuming the initial orbit was circular). We assumed in these
calculations that the B~stars have a mass of $15\,\MO$, as corresponds
to a B0-1 main-sequence star, and that the neutron star has a mass of
$1.4\,\MO$. (For B-star masses in the range $10-20\,\MO$ the required
minimum runaway velocities do not differ by more than $\pm$ 10 per
cent from the values for $15\MO$). The table shows that the required
minimum kick velocities range from about 50\,\kms\, to about 200
\kms. Assuming the real kick velocities to be randomly distributed,
the required kicks become $\sqrt{3/2}$ times larger, and range from
about 60 to about 250 \kms.

We conclude from the above that the combination of low mean space
velocity of the Be/X-ray binaries and large mean orbital eccentricity
provides unequivocal evidence for the existence of velocity kicks
imparted to neutron stars at their birth.

An alternative way to approach the problem of the orbital
eccentricities is to calculate, from the measured mean runaway
velocities of Be/X-ray systems, what orbital eccentricity these
systems should have had, were this runaway velocity imparted by purely
symmetric mass ejection. This is the topic of the next section.

\begin{table*}
\caption[]{Orbital parametres of the seven Be/X-ray binaries with
known orbital periods and measured or estimated orbital
eccentricities, together with those of the two radio pulsars with
B-type companions. Column 5 gives an estimate of the recoil velocity
of the binary assuming symmetric mass loss.  Columns 6 and 7 indicate
the orbital eccentricities which would have been induced by symmetric
supernova mass ejection that imparted runaway velocities of 10 and 20
\kms\, to the systems, respectively. Column 8 lists the minimum kick
velocities, that should have been imparted to their neutron stars to
produce their actually observed orbital eccentricities without
imparting a runaway velocity larger than 20 \kms\, to the
systems. References to the orbital parameters of the systems: (1) van
Paradijs (1985); (2) Corbet at al. (1986); (3) Priedhorsky \& Terrell
(1983); (4,5) Parmar et al. (1989a, 1989b); (6) Johnston et
al. (1992); (7) Kaspi et al. (1994); (8) Kaspi et al. (1996a); (9)
Kaspi et al. (1996b).  }
\begin{minipage}{18cm}
\renewcommand{\footnoterule}{}
\begin{tabular}{lrcrrccrr}
\hline\noalign{\smallskip}
\multicolumn{1}{c}{System}                &  
\multicolumn{1}{c}{$P_{\rm orb}$}         & 
\multicolumn{1}{c}{$e$}                   & 
\multicolumn{1}{c}{$\langle V_{\rm orb} \rangle$\footnote{As defined
                                                          by Eq.~(20)}}&
\multicolumn{1}{c}{$V_{\rm rec}$\footnote{From Eq.\,(22)}}   &  
\multicolumn{2}{c}{`symmetric' $e$-expected for}           & 
\multicolumn{1}{c}{required}              & 
\multicolumn{1}{c}{Ref.}                  \\
\noalign{\smallskip}
\multicolumn{1}{c}{ }                     & 
\multicolumn{1}{c}{[days]}                & 
\multicolumn{1}{c}{(observed)}            & 
\multicolumn{1}{c}{[\kms]}                &  
\multicolumn{1}{c}{[\kms]}                &  
\multicolumn{1}{c}{$V_{\rm rec}=10$}      &  
\multicolumn{1}{c}{$V_{\rm rec}=20$}      &
\multicolumn{1}{c}{minimum}              &
\multicolumn{1}{c}{ }                     \\
\noalign{\smallskip}             
\multicolumn{1}{c}{ }                     &
\multicolumn{1}{c}{ }                     &  
\multicolumn{1}{c}{ }                     &
\multicolumn{1}{c}{ }                     &
\multicolumn{1}{c}{ }                     &
\multicolumn{1}{c}{[\kms]}                  & 
\multicolumn{1}{c}{[\kms]}                  &   
\multicolumn{1}{c}{$V_{\rm kick}$ [\kms]} & 
\multicolumn{1}{c}{ }                     \\ 
\hline\noalign{\smallskip}
X0115+634       &  24.3\hspace{2mm}  & \hspace{4mm} 0.34       &60.1   & 182.8 & 0.06 & 0.11 &  66\hspace{8mm}   & (1)   \\
X0331$+$53      &  34.3\hspace{2mm}  & \hspace{4mm} 0.31       &53.5   & 162.7 & 0.06 & 0.12 &  53\hspace{8mm}   & (1,3) \\ 
X0535$+$26      & 111.0\hspace{2mm}  & $\geq 0.4$              &42.8   & 108.4 & 0.09 & 0.18 &  47\hspace{8mm}   & (1)   \\
X0535$-$67      &  16.7\hspace{2mm}  & $\geq 0.7$              &184    & 207\hspace{2mm}   & 0.05 & 0.10 & 203\hspace{8mm}   & (1)   \\
X1145$-$619     & 187.5\hspace{2mm}  & $\geq 0.7\footnote{Estimated in the text}$        
                                                               &82.3    &  92.4 & 0.11 & 0.21 &  90.5\hspace{6mm} & (1)   \\
X1258$-$613     & 133\hspace{4mm}    & $\geq 0.7^{c}$          &92.7    & 103.9 & 0.10 & 0.19 & 102.0\hspace{6mm} & (2,3) \\
EXO2030$+$375    &  46\hspace{4mm}    & \hspace{4mm} 0.38      &55.1    & 147.6 & 0.07 & 0.13 &  60.6\hspace{6mm} & (4,5) \\ 
\hline\noalign{\smallskip}
PSRJ1259$-$63   & 1236.8\hspace{2mm} &\hspace{4mm} 0.87       &79.1    & 49.3  & 0.20 & 0.39 & 87\hspace{8mm}    & (6) \\
PSRJ0045$-$7319 & 51.17              &\hspace{4mm} 0.81       &179     & 142.5 & 0.07 & 0.14 & 195.4\hspace{6mm} & (7,8,9) \\
\hline\noalign{\smallskip}
\end{tabular}
\label{Tab:observed_v}
\end{minipage}
\end{table*}

\subsection{Predicted relation between orbital eccentricity and
runaway velocity expected in case of symmetric explosions - comparison
with observations} 
    
Eq.~(14) yields:
\begin{equation}
          {e \over 1 + e} = {\DMsn \over \Mf},
\label{Eq:17}\end{equation}
Combination of Eqs~(\ref{Eq:11}) and (\ref{Eq:17}) yields:
\begin{equation}
\Vrec = \sqrt{ {G\Mfsn \over a'} } \frac{\Msf}{\Mfsn} 
	            \frac{e}{1+e},
\label{Eq:18}\end{equation} 
where $a'$ is the pre-supernova orbital radius.  The semi-major axis
after the supernova $a''$ follows from:
\begin{equation}
{a' \over a''} = 1 - {\DMsn \over \Mfsn}, 
\end{equation}
and by writing
\begin{equation}
\Mf = \Mfsn+\DMsn = \Mfsn(1 + {\DMsn \over \Mfsn}).
\label{Eq:20}\end{equation}
one obtains after insertion of
Eq.~(18) in Eq.~(16):
\begin{equation}
\Vrec^2 = \frac{G\Mfsn}{a''}
	\frac{1+\DMsn/\Mfsn}{1-\DMsn/\Mfsn}
	\left(\frac{\Msf}{\Mfsn}                \right)^2
	\left(\frac{e}{1+e}               \right)^2.
\label{Eq:21}\end{equation}

Defining now the presently observed mean orbital velocity by
\begin{equation}
         \langle \Vorb \rangle^2 = {G\Mfsn \over a''},
\label{Eq:22}\end{equation}
and substituting $\DMsn/\Mfsn$\, from Eq.~(\ref{Eq:13}) one obtains:
\begin{equation}
	{\Vrec \over \langle \Vorb \rangle}
        {\Mfsn \over \Msf} = \frac{e}{(1-e^2)^{1/2}}.
\label{Eq:23}\end{equation} 
This defines, in the case of symmetric supernova-mass ejection the
relation that is expected to be found between the observed system
runaway velocity \Vrec\ and the observed orbital eccentricity $e$, for
a system with a Be/X-ray star of mass \Msf, and observed mean orbital
velocity $\langle \Vorb \rangle$.  Since $\Mfsn = \Msf+1.4$\,\MO, and
since in general $\Msf >10$\,\MO, the quantity $\Mfsn/\Msf$ is close
to unity.  Defining:
\begin{equation} 
	f_v \equiv \frac{\Vrec}{\Vorb} \, {\Mfsn \over \Msf} = \frac {e}{(1-e^{2})^{1/2}},
\label{Eq:24}\end{equation} 
one obtains a simple relation between $f_v$ and $e$, the plotted curve
in Fig.~2. In the case of symmetric supernova mass ejection, the
observed value of $f_v$ of a Be/X-ray binary should be related to the
observed orbital eccentricity according to this curve, which shows
that large eccentricities correspond to large runaway velocities.

\begin{figure}[t]
\centerline{\psfig{file=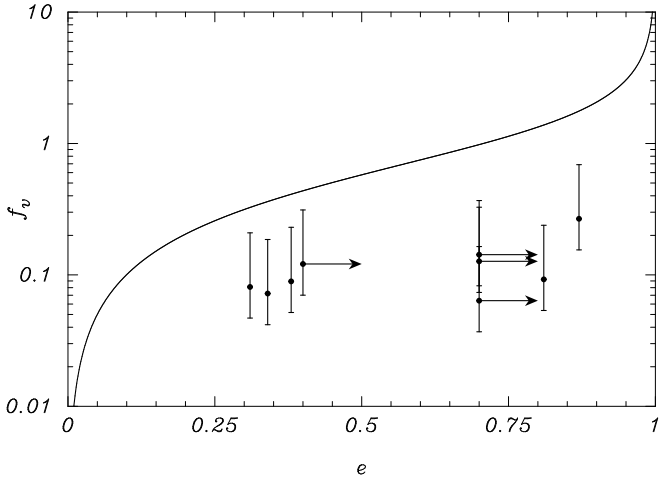,height=7.5cm,angle=-90}}
\caption[]{The system velocity of the runaway binary as a function of
the orbital eccentricity induced upon the symmetric explosion of the
core of the primary star. Along the vertical axis is the expression:
$f_v \equiv \Vrec/\Vorb \times \Mfsn/\Msf$ (which is a dimension-less
quantity). The observed positions of the nine Be/neutron star systems
in this diagram show that the explosions cannot have been symmetric.}
\label{Fig:Vfrac}
\end{figure}

In Fig.~2 we also plotted the values of $f_{v}$ and $e$ for the nine
systems with observed orbital periods and eccentricities (see
Table~3), taking runaway velocities $V_{\rm rec}$ in the observed
range $19 \pm 8$ \kms\, for the Be X-ray binaries. We assumed a
Be-star mass of $15\MO$. The figure shows that all systems fall far
below the curve expected for symmetric supernova mass ejection. This
again shows that the combination of low runaway velocities and large
orbital eccentricities as observed in the Be/X-ray binaries cannot be
obtained by symmetric mass ejection in the supernovae, and that a
velocity kick imparted to the neutron stars at birth is absolutely
required.

\section{Conclusions}

The measured tangential velocities of the Be/X-ray binaries and
OB-supergiant X-ray binaries by the {\it Hipparcos} satellite confirm
the expectations from the evolution of massive close binaries in which
little mass is lost from the binary systems during the first mass
transfer phase. The much higher tangential velocities of supergiant
X-ray binaries than those of the Be-systems follow from a combination
of (1) the much larger fractional helium core masses in the
progenitors of the OB-supergiant systems which cause their
pre-supernova orbital periods to be shorter, and thus their
pre-supernova orbital velocities to be much larger than those of the
less massive Be-systems, and (2) the much lower amounts of mass
ejected during the supernova explosion in the lower-mass Be-systems
compared to the OB-supergiant systems.

The combination of a high orbital eccentricity with a low space
velocity observed for the Be type X-ray binaries can only be
understood if a kick with appreciable velocity --in the range 60 to
250~\kms-- is imparted to the newly born neutron star. Such a kick
tends to only slightly affect the space velocity of the binary system
since the neutron star has to drag along its massive companion. The
orbital eccentricity, however, is strongly affected by such a
asymmetric velocity kick. If the supernova explosions in these systems
had been symmetric, the high orbital eccentricities observed in the
class of Be X-ray binaries are impossible to reconcile with their on
average low runaway velocities.

\acknowledgements{This work was supported by Spinoza Grant 08-0 to
E.P.J. van den Heuvel and by NASA through Hubble Fellowship grant
HF-01112.01-98A awarded (to SPZ) by the Space Telescope Science
Institute, which is operated by the Association of Universities for
Research in Astronomy, Inc., for NASA under contract NAS\, 5-26555. LK
is supported by a fellowship of the Royal Netherlands Academy of Arts
and Sciences.  The first two authors thank the Institute for
Theoretical Physics UCSB, where most of this research was carried out
in the fall of 1997. 
}


\begin{thebibliography}{}
\bibitem{} Chevalier C. Ilovaisky, S., A.\ 1998, A\&A 330, 201
\bibitem{} Clark, L.L., Dolan, J.F. 1999, A\&A 350, 1085
\bibitem{} Comer\'{o}n, F., Torra, J., G\'{o}mez, A.E. 1998 A\&A 330, 975
\bibitem{} Corbet, R.H.D., Smale, A.P., Menzies, J.W., Branduardi Raymont, G., Charles, P.A., Mason, K.O., Booth, L.\ 1986, MNRAS 221, 961
\bibitem{} Crampton, D., Hutchings, J.B., Cowley, A.P.\ 1978 ApJ
225, L63 
A\&A 142, 367
\bibitem{} de Loore, C., De Greve, J.\, P. 1992, A\&ASS 94, 453
\bibitem{} Ergma, E., van den Heuvel, E.\ P.\ J.\ 1998 A\&A 331, L29
\bibitem{} Flannery, B. P., Ulrich, R. K., 1977, ApJ 212, 533
\bibitem{} Gies, D.R., Bolton, C.T. 1986, ApJS 61, 419
\bibitem{} Haberl, F. 1995, A\&A 296, 685
\bibitem{} Habets, G.\ M.\ H.\ J.\ 1985, PhD Thesis, U. Amsterdam
\bibitem{} Habets, G.\ M.\ H.\ J.\ 1986a, A\&A 165, 95 
\bibitem{} Habets, G.\ M.\ H.\ J.\ 1986b, A\&A 167, 61
\bibitem{} Hartman, J. W., 1997, A\&A 322, 127
\bibitem{} Hills, J.\ 1983, ApJ 267, 322
\bibitem{} Iben I.\ J., Tutukov, A.\ V.\ 1985, ApJS 58, 661
\bibitem{} Iben I.\ J., Tutukov, A.\ V.\ 1998, ApJ 501, 263
\bibitem{} Johnston, J., Manchester, R.R., Lyme, A.G., Bailes, M., Kaspi, V.M., Guojun, Quand, d`Amico, N.\ 1992, ApJ 387, L37
\bibitem{} Kaper, L., Comer\'{o}n, F., Barziv, O. 1999, in Proc.\ IAU
Symp. 193, p. 316
\bibitem{} Kaspi, V. Johnston, S., Bell, J.F, Manchester, R.N., Bailes, M., Bessel, M., Lyne, A.G., d`Amico, N.\ 1994, ApJ 243, L43
\bibitem{} Kaspi, V.M., Tauris, T.M., Manchester, R.N.\ 1996a, ApJ 459, 717
\bibitem{} Kaspi, V.M., Bailes, M., Manchester, R.N, Stappers, B.V., Bell, J.F.\ 1996b, Nat 381, 584
\bibitem{} Kochanek, C., S.\ 1992, ApJ 385, 604 
\bibitem{} Leitherer, C., Chapman, J.M., Koribalski, B. 1995 ApJ 450,
289 
\bibitem{} Moffat, A.F.J., Marchenko, S.V., Van der Hucht, K.A., et
al. 1998, A\&A 331, 949
\bibitem{} Paczy\'nski, B. 1971, Acta.\ Astron.\ 21, 417
\bibitem{} Paczy\'nski, B. 1990 ApJ 348, 485
\bibitem{} Parmar, A. N., White, N. E., Stella, L., Izzo, C., Ferri, P. 1989, ApJ 338, 359 
\bibitem{} Parmar, A. N., White, N. E., Stella, L. 1989, ApJ 338, 373
\bibitem{} Portegies~Zwart, S.F., 1995 A\&A 296, 691
\bibitem{} Portegies~Zwart, S.F., \& van den Heuvel, E.P.J. 1999, New
Astron 4, 355
\bibitem{} Portegies~Zwart, S.F., 2000 ApJ {\em in press}
(astro-ph/0005021)
\bibitem{} Priedhorsky, W.C., Terrel, J.\ 1983, ApJ 273, 709
\bibitem{} Steele, I. A., Negueruela, I., Coe, M. J., Roche, P. 1998,
MNRAS 297, L5 
\bibitem{} Soberman, G. E., Phinney, E., S., van den Heuvel, E. P. J.,
1997, A\&A 327, 620
\bibitem{} van den Heuvel, E. P. J.\ 1983, in: Accretion-driven
stellar X-ray sources P.\ 308 
\bibitem{} van den Heuvel, E. P. J.\ 1985, in: Birth and evolution of
massive stars and stellar groups; Proceedings of the Symposium,
Dwingeloo, Netherlands, (Dordrecht, D. Reidel) p.\ 107
\bibitem{} van~den Heuvel, E. P. J. 1994, A\&A 291, L39 
\bibitem{} van den Heuvel, E. P. J.,  Rappaport, S.\ 1987,
in: Physics of Be stars; Proceedings of the Ninety-second IAU
Colloquium, (Cambridge University Press, 1987) p.\ 291 
\bibitem{} van den Linden, Th.\ 1982, Ph. Thesis, University of Amsterdam
\bibitem{} van Oijen, J.G.J.\ 1989, A\&A 217, 115
\bibitem{} van Paradijs, J.\ 1995, in: "X-ray Binaries",
(eds. W.H.G. Lewin, J.A. van Paradijs and E.P.J. van den Heuvel), Cambridge University Press, 536-577
\bibitem{} Wellstein, S., Langer, N. 1999, A\&A 350, 148
\bibitem{} Woosley, S. E., Langer, N., Weaver, T. A. 1995, ApJ 448, 315
\bibitem{} Wijers, R. A. M. J., van Paradijs, J., van den Heuvel, E. P. J. 1992, A\&A 261, 145
\bibitem{} Zahn, J., P.\ 1977, A\&A 57, 383
\end{thebibliography}
\end{document}